\documentclass[manuscript]{emulateapj}

\usepackage{psfig,graphics} 

\def\subsun{\mbox{$_{\odot}$}}

\shorttitle{Three Modes of Star Formation}
\shortauthors{Smith, Turk, Sigurdsson, O'Shea, \& Norman}

\begin{document}

\title{Three Modes of Metal-Enriched Star Formation in the Early Universe}

\shorttitle{Three Modes of Metal-Enriched Star Formation}

\author{Britton D. Smith\altaffilmark{1}}
\affil{525 Davey Laboratory,
  Department of Astronomy \& Astrophysics,
  The Pennsylvania State University,
  University Park, PA 16802}
\email{brittons@origins.colorado.edu}

\author{Matthew J. Turk}
\affil{Kavli Institute for Particle Astrophysics and Cosmology, 
  2575 Sand Hill Rd., Mail Stop 29, 
  Menlo Park, CA 94025}
\email{mturk@slac.stanford.edu}

\author{Steinn Sigurdsson}
\affil{525 Davey Laboratory, 
  Department of Astronomy \& Astrophysics, 
  The Pennsylvania State University, 
  University Park, PA 16802}
\email{steinn@astro.psu.edu}

\author{Brian W. O'Shea\altaffilmark{2}}
\affil{Department of Physics \& Astronomy,
  Michigan State University,
  East Lansing, MI 48824}
\email{oshea@msu.edu}

\and

\author{Michael L. Norman}
\affil{Center for Astrophysics and Space Sciences,
  University of California at San Diego, La Jolla, CA 92093}
\email{mlnorman@ucsd.edu}

\altaffiltext{1}{Center for Astrophysics \& Space Astronomy,
  Department of Astrophysical \& Planetary Sciences,
  University of Colorado, Boulder, CO, 80309}
\altaffiltext{2}{Theoretical Astrophysics Group,
  Los Alamos National Laboratory,
  Los Alamos, NM 87545}

\begin{abstract}
Simulations of the formation of Population III (Pop III) stars suggest that they 
were much more massive than the Pop II and Pop I stars observed today. This 
is due to the collapse dynamics of metal-free gas, which is regulated by the 
radiative cooling of molecular hydrogen. We study how the collapse 
of gas clouds is altered by the addition of metals to the star-forming 
environment by performing a series of simulations of 
pre-enriched star formation at various metallicities.  To make a clean 
comparison to metal-free star formation, we use initial conditions identical 
to a Pop III star formation simulation, with low ionization and 
no external radiation other than the cosmic microwave background (CMB).  
For metallicities below the critical metallicity, $Z_{cr}$, 
collapse proceeds similarly to the metal-free case, and only massive objects form.  
For metallicities well above $Z_{cr}$, efficient cooling rapidly lowers the gas 
temperature to the temperature of the CMB.  The gas is 
unable to radiatively cool below the CMB temperature, and becomes 
thermally stable.  For high metallicities, $Z \ga 10^{-2.5} Z\subsun$, 
this occurs early in the evolution of the gas cloud, when the 
density is still relatively low.  The resulting cloud-cores show little or 
no fragmentation, and would most likely form massive stars.  If the metallicity is not vastly 
above $Z_{cr}$, the cloud cools efficiently but does not reach the CMB temperature, and 
fragmentation into multiple objects occurs.  We conclude that there were three distinct modes of star 
formation at high redshift ($z \ga 4$): a `primordial' mode, producing massive stars 
(10s to 100s $M\subsun$) at very low metallicities ($Z \la 10^{-3.75} Z\subsun$); a CMB-regulated mode, 
producing moderate mass (10s of $M\subsun$) stars at high metallicites ($Z \ga 10^{-2.5} Z\subsun$ at 
redshift $z\sim$15-20); and a low-mass (a few $M\subsun$) mode existing between those two metallicities.  
As the universe ages and the CMB temperature decreases, the range of the low mass mode extends to higher 
metallicities, eventually becoming the only mode of star formation.
\end{abstract}

\keywords{stars: formation, cosmology}

\section{Introduction}

Understanding the nature of the first stars in the universe is, in principle, a very 
straightforward problem to solve.  The initial conditions are defined by well-constrained 
cosmological parameters (e.g., \citet{2007ApJS..170..377S,2008arXiv0803.0547K}) and accepted 
calculations of 
Big Bang Nucleosynthesis (e.g., \citet{1993ApJS...85..219S}).  The relative simplicity of 
the chemistry of primordial gas \citep{1997NewA....2..181A,1998A&A...335..403G}, along with 
powerful numerical methods, have made it possible to accurately characterize the formation 
process of primordial stars from the assembly of their dark matter halos through the end of 
the optically-thin regime of the collapsing protostar \citep{2002Sci...295...93A,2002ApJ...564...23B,
2004NewA....9..353B,2006ApJ...652....6Y,2007ApJ...654...66O,2007MNRAS.378..449G}.  Results 
from theory and numerical simulations suggest that the first stars were 10s to 
100s $M\subsun$ \citep{2002Sci...295...93A,2003ApJ...589..677O,2004ApJ...603..383T,
2006ApJ...652....6Y,2007ApJ...654...66O}.  
A direct calculation, however, involves simulating the complex processes of accretion and 
radiative feedback, a capability currently just beyond the state of the art.  
In contrast, the challenge of understanding the second generation of stars is significantly more 
complex.  The initial conditions of the first stars essentially depend only on the 
principles of cold dark matter cosmology and the properties of molecular hydrogen, 
but the initial conditions of the second stars 
require a complete solution to the formation and evolution of their predecessors as well as host of 
additional chemistry and physical processes.  The 
chemical composition and physical conditions of second-generation star-forming environments 
strongly depend on the exact masses of the first stars \citep{2002ApJ...567..532H,
2005ASPC..336...79M,2006NuPhA.777..424N,2006astro.ph..8028R} and the mechanics of their 
supernovae \citep{2005ApJ...630..675K,2007ApJ...670....1G}.

At the heart of the problem regarding forming the second stars is understanding how the star 
formation 
process is altered by the introduction of the first metals in the universe, created in the 
explosions of the first stars.    
The chemical composition of a gas determines the efficiency with which it can cool radiatively.  
Cold, metal-free gas ($T < 10^{4}$ K) cools solely through ro-vibrational lines of H$_{2}$, whose 
lowest-lying transition has an energy equivalent temperature of $\sim$ 512 K, resulting in 
a minimum achievable temperature of $\sim$ 200 K.  The 
energy levels of H$_{2}$ become thermalized at rather low densities, $n$ $\sim$ 10$^{3-4}$ 
cm$^{-3}$, where $n$ is the number density, above which the cooling rate is only proportional to 
$n$, instead of $n^{2}$.  Numerical 
simulations have shown that this creates a stalling point in the star formation process, 
where the free-fall collapse of metal-free gas comes to a halt as the cooling time increases to  
be significantly above the dynamical time \citep{2002Sci...295...93A,2002ApJ...564...23B}.  Entry 
into this `loitering' phase marks the end of hierarchical fragmentation that occurs 
during the free-fall period, a fragmentation that occurs because the 
temperature is able to continually decrease with 
increasing density.  The final fragmentation mass scale is then set by the Jeans mass 
at the temperature and density corresponding to the point where the gas can no longer get colder 
with increasing density \citep{1985MNRAS.214..379L,2005MNRAS.359..211L}.  
In the metal-free case, this yields a mass scale of $\sim$ 1000 $M\subsun$, resulting in the 
high-mass nature of the first stars.

The addition of metals enhances the cooling rate at low temperatures through fine-structure 
and molecular transitions, as well as by continuum emission from dust grains.  At low to moderate 
densities ($n <$ 10$^{9}$ cm$^{-3}$), only metals in the gas phase contribute significantly to 
the cooling, unless the level of metal enrichment is very high \citep{2008MNRAS.385.1443S}.  
In the phases where H$_{2}$ cooling is strong ($n \la 10^{4}$ cm$^{-3}$, an increase in the cooling 
due to metals is first seen only at temperatures less 
than the 200 K temperature minimum that exists in metal-free star formation 
\citep{2008MNRAS.385.1443S}.  At abundances of $Z \la$ 10$^{-4}$ $Z\subsun$, where metal cooling is 
only significant for $T <$ 200 K, the low-density evolution of the cloud is identical 
to the metal-free case \citep{2000ApJ...534..809O,2001MNRAS.328..969B,2005ApJ...626..627O}.  
At higher metallicities, gas-phase metal cooling becomes strong enough for the collapsing cloud to 
bypass the loitering phase and to undergo continued fragmentation 
\citep{2001MNRAS.328..969B,2003Natur.425..812B,2006ApJ...643...26S,2007ApJ...661L...5S}.  
If dust grains are present, their influence becomes important at very high densities 
($n >$ 10$^{12}$ cm$^{-3}$).  This has been shown to induce fragmentation for metallicities 
as low as 10$^{-5.5} Z\subsun$ \citep{2005ApJ...626..627O,2006MNRAS.369.1437S,2006ApJ...642L..61T,
2008ApJ...672..757C}.  Calculations by \citet{2004MNRAS.351.1379S} predict that up to 30\% of the 
progenitor mass is converted into dust in a pair-instability supernova, but observations of Type II 
supernova in the local universe remnants have not returned conclusive evidence of dust 
\citep{2004MNRAS.355.1315G,2004Natur.432..596K}.  
In the context of star formation in the early universe, the arrival at 
this critical metallicity, $Z_{cr}$, at which fragmentation occurs beyond the capabilities of 
primordial gas, is predicted to be the point where the universal mode of star formation shifts 
permanently from the high-mass, solitary mode of the first stars, to the low-mass, 
multiply-producing mode that is presently observed  \citep{2003Natur.425..812B,2006ApJ...643...26S,
2007ApJ...661L...5S}.

\begin{deluxetable*}{lccccccc}
  \tablecolumns{8}
  \tablewidth{0pt}
  \tablecaption{Simulations}
  \tablehead{
    \colhead{Run} &
    \colhead{IC} &
    \colhead{$Z$ (Z$\subsun$)} &
    \colhead{$z_{col}$} &
    \colhead{Grids} &
    \colhead{Cells ($\times10^{7}$)} &
    \colhead{$n_{max}$ (cm$^{-3}$)} &
    \colhead{$\Delta t_{col}$ (yr)}}
  \startdata
  r1\_mf       & 1 & 0            & 14.761 & 13790 & 6.49 & 5.96 $\times 10^{11}$ & -\\
  r1\_Z-6      & 1 & 10$^{-6}$     & 14.762 & 13015 & 6.42 & 6.43 $\times 10^{11}$ & 2.9 $\times10^{4}$\\
  r1\_Z-5      & 1 & 10$^{-5}$     & 14.783 & 12889 & 6.43 & 6.74 $\times 10^{11}$ & 5.7 $\times10^{5}$\\
  r1\_Z-4.25   & 1 & 10$^{-4.25}$  & 14.809 & 12955 & 6.40 & 6.75 $\times 10^{11}$ & 1.2 $\times10^{6}$\\
  r1\_Z-4      & 1 & 10$^{-4}$     & 14.830 & 12964 & 6.38 & 6.22 $\times 10^{11}$ & 1.8 $\times10^{6}$\\
  r1\_Z-3.75   & 1 & 10$^{-3.75}$  & 14.848 & 13003 & 6.38 & 6.97 $\times 10^{11}$ & 2.2 $\times10^{6}$\\
  r1\_Z-3.5    & 1 & 10$^{-3.5}$   & 14.874 & 12883 & 6.40 & 6.89 $\times 10^{11}$ & 2.9 $\times10^{6}$\\
  r1\_Z-3.25   & 1 & 10$^{-3.25}$  & 14.936 & 12888 & 6.36 & 3.97 $\times 10^{11}$ & 4.5 $\times10^{4}$\\
  r1\_Z-3      & 1 & 10$^{-3}$     & 15.073 & 12581 & 6.25 & 4.42 $\times 10^{11}$ & 7.9 $\times10^{6}$\\
  r1\_Z-2.5    & 1 & 10$^{-2.5}$   & 16.180 & 11187 & 5.69 & 4.70 $\times 10^{11}$ & 3.3 $\times10^{7}$\\
  r1\_Z-2      & 1 & 10$^{-2}$     & 19.481 & 8693  & 4.57 & 1.56 $\times 10^{11}$ & 8.8 $\times10^{7}$\\
  \tableline
  r2\_mf       & 2 & 0            & 17.409 & 8684  & 4.76 & 7.97 $\times 10^{11}$ & -\\
  r2\_Z-4      & 2 & 10$^{-4}$     & 17.555 & 8509  & 4.74 & 8.39 $\times 10^{11}$ & 2.5 $\times10^{6}$\\
  r2\_Z-3.5    & 2 & 10$^{-3.5}$   & 17.654 & 8476  & 4.71 & 8.20 $\times 10^{11}$ & 4.2 $\times10^{6}$\\
  r2\_Z-3      & 2 & 10$^{-3}$     & 17.955 & 8408  & 4.67 & 7.30 $\times 10^{11}$ & 9.2 $\times10^{6}$\\
  r2\_Z-2.5    & 2 & 10$^{-2.5}$   & 18.537 & 8022  & 4.51 & 4.41 $\times 10^{11}$ & 1.8 $\times10^{7}$\\
  r2\_Z-2      & 2 & 10$^{-2}$     & 20.441 & 7194  & 4.09 & 1.92 $\times 10^{11}$ & 4.4 $\times10^{7}$\\
  \tableline
  r3\_mf       & 3 & 0            & 23.885 & 7771  & 4.25 & 1.63 $\times 10^{12}$ & -\\
  r3\_Z-4      & 3 & 10$^{-4}$    & 23.966 & 7722  & 4.23 & 1.65 $\times 10^{12}$ & 6.7 $\times10^{5}$\\
  r3\_Z-3.5    & 3 & 10$^{-3.5}$  & 24.122 & 7640  & 4.21 & 1.42 $\times 10^{12}$ & 1.9 $\times10^{6}$\\
  r3\_Z-3      & 3 & 10$^{-3}$    & 24.390 & 7366  & 4.17 & 1.29 $\times 10^{12}$ & 4.1 $\times10^{6}$\\
  r3\_Z-2.5    & 3 & 10$^{-2.5}$  & 24.732 & 7145  & 4.13 & 1.20 $\times 10^{12}$ & 6.7 $\times10^{6}$\\
  r3\_Z-2      & 3 & 10$^{-2}$    & 25.028 & 7083  & 4.09 & 6.53 $\times 10^{11}$ & 8.9 $\times10^{6}$\\
  r3\_Z-2noCMB & 3 & 10$^{-2}$    & 25.255 & 7424  & 4.29 & 6.02 $\times 10^{11}$ & 1.1 $\times10^{7}$\\
\enddata
\tablecomments{$z_{col}$ is the redshift at the onset of runaway collapse.  The total number of grid cells includes 
those that are covered by child grids at higher levels of refinement.  $n_{max}$ is the proper maximum baryon number 
density within the box.  $\Delta$t$_{col}$ is the time difference to runaway collapse from the metal-free case.}
\label{tab:sims}
\end{deluxetable*}

The protostellar collapse of metal-enriched gas clouds has been studied extensively with 
one-zone models coupled to large chemical networks \citep{2000ApJ...534..809O,
2003Natur.422..869S,2005ApJ...626..627O,2006MNRAS.369.1437S}.  These studies have 
produced insight into the evolution of the density and temperature of collapsing gas clouds 
with finite metallicities, but one-zone models cannot speak to the actual process of fragmentation, 
which requires attention to complex cloud geometries that can only be given by fully  
three-dimensional hydrodynamic simulations \citep{2007astro.ph..1733L}.  The first of such 
simulations were carried out by \citet{2001MNRAS.328..969B}, who included cooling from 
C, N, O, Fe, S, and Si, but not H$_{2}$, finding that clouds with metallicities, $Z$ $\ge$ 
10$^{-3}$ $Z\subsun$ are able to fragment to mass scales lower than in the metal-free 
case.  These simulations had a mass resolution of only 100 $M\subsun$ and were, therefore, 
unable to investigate the formation of solar-mass stars.  More recently, 
\citet{2007ApJ...661L...5S} have performed a series of high-resolution simulations of 
metal-enriched gas collapse that were able to follow the evolution of gas fragments to 
sub-solar mass scales.  They, too, find that gas with metallicities, $Z$ $\ge$ 10$^{-3}$ 
$Z\subsun$ will fragment into multiple clumps, while gas with $Z$ $\le$ 10$^{-4}$ $Z\subsun$ 
will produce only one object.  However, \citet{2008MNRAS.385.1443S} reported the 
existence of regions in density and temperature that are thermally unstable in gas 
with metallicities as low as 10$^{-4}$ $Z\subsun$.  Fragmentation is traditionally thought 
to happen when the cooling time is less than the dynamical time, as the gas is able to 
cool and form perturbations before they can be smoothed out by sound waves.  When this 
condition, referred to as the fragmentation criterion, is satisfied, fragmentation 
can also be aided by thermal instabilities \citep{1965ApJ...142..531F}, where a slight 
decrease in temperature or increase in density leads to a higher cooling rate, causing 
differences in temperature between pockets of gas to grow in a runaway fashion.  
In the adaptive mesh refinement 
simulations of \citet{2007ApJ...661L...5S}, grid refinement was performed based on baryon 
and dark matter overdensities and by ensuring that the Truelove criterion, $l_{J}$ $<$ $\Delta x$, 
where $l_{J}$ is the local Jeans length and $\Delta x$ is the grid cell size, was satisfied 
\citep{1997ApJ...489L.179T}.  However, refinement was not performed when the cooling time was less than 
the hydrodynamic time-step.  While this was not explicitly wrong, since the radiative cooling 
solver iterates with time-steps that are no larger than 10\% of the cooling time, it 
may have artificially suppressed the growth of thermal instabilities that could have formed extra 
fragments.

We rerun the simulations of \citet{2007ApJ...661L...5S} with an additional refinement criterion 
which ensures that the hydrodynamic time-step is always less than the cooling time on the finest 
level of resolution.  
In addition, we extend the series of simulations to include lower metallicities and to more carefully 
examine the metallicity range 
between 10$^{-4}$ $Z\subsun$ and 10$^{-3}$ $Z\subsun$.  We also run similiar simulations with 
two extra sets of initial conditions to confirm the robustness of the results.  
We describe the setup of our simulations and the improvements over \citet{2007ApJ...661L...5S} in 
\S\ref{sec:setup}.  In \S\ref{sec:results}, we calculate the gas-phase critical metallicity, evaluate 
the validity of our assumption of optical thinness, and present the results of the suite of simulations.  
In \S\ref{sec:discussion}, we discuss our results in the context of star formation at high redshift and 
make the case for an initial mass function (IMF) that evolves over cosmic time.  We also include a 
discussion of the caveats and limitations of this work.  Finally, we conclude with a brief summary of the 
main conclusions of this work in \S\ref{sec:conclusion}.

\section{Simulation Setup} \label{sec:setup}

We perform a series of 24 primordial star formation simulations using the Eulerian adaptive mesh 
refinement hydrodynamics + N-body code, Enzo \citep{1997WSAMRGMBryan,2004CWAMROShea}.  Excluding 
our three metal-free control runs, the gas in each simulation is homogeneously pre-enriched to some 
non-zero metallicity.  As in \citet{2007ApJ...661L...5S}, we confine the simulations to constant 
metallicities with solar abundance patterns, saving the more realistic, and far more complicated, 
simulations of true 
second-generation star-forming environments, with heterogeneous metal-mixing and nonsolar abundance 
patterns, for a future work.

The nature of the initial conditions for our simulations are identical to those used in 
\citet{2007ApJ...661L...5S}.  The simulation box has a comoving size of 300 $h^{-1}$ kpc, 
with 128$^{3}$ grid cells on the top grid and three nested subgrids, each refining by a 
factor of 2, for an effective top grid resolution of 1024$^{3}$ cells.  The cosmological 
parameters have the following values:  $\Omega_{M}$ = 0.3, $\Omega_{\Lambda}$ = 0.7, 
$\Omega_{B}$ = 0.04, and Hubble constant, $h$ = 0.7, in units of 100 km s$^{-1}$ Mpc$^{-1}$.  
The power spectrum of initial density fluctuations is given by \citet{1999ApJ...511....5E}, 
with $\sigma_{8}$ = 0.9 and $n$ = 1.  Refined grids are created during the simulations when 
the baryon (dark matter) density is 4 (8) times greater than the mean density at that level.  
The density threshold for refinement decreases at higher levels.  
The local Jeans length is resolved by a minimum of 16 grid cells at all times, exceeding the 
Truelove criterion \citep{1997ApJ...489L.179T} by a factor of four along each coordinate axis.  
In addition, grid refinement occurs 
whenever the cooling time drops below the integration time step of the hydrodynamic solver.  
This final refinement criterion was not used in \citet{2007ApJ...661L...5S}.  During the simulation, 
a grid cell is flagged for refinement if one or more of any of these criteria are met.

We perform three sets of simulations.  Qualitatively, the three sets are the same.  They each 
have the same cosmological parameters, box size, and resolution.  The only difference between 
them is that their initial conditions were created with three unique randomizations of the initial 
density and velocity perturbations.  Thus, they represent three different realizations of the same 
problem.  The first set of initial conditions is the one used by \citet{2007ApJ...661L...5S}.  
The second and third sets were initially used by \citet{2007ApJ...654...66O} and correspond to the 
runs named L0\_30A and L0\_30D in that work.  For each set of initial conditions, we perform a 
metal-free control run.  Excluding the control runs, we run 10 simulations using initial condtions 
Set 1, with metallicities ranging from 10$^{-6} Z\subsun$ to 10$^{-2} Z\subsun$, 5 simulations 
using Set 2, and 6 simulations using Set 3, with metallicities from 10$^{-4} Z\subsun$ to 
10$^{-2} Z\subsun$ for Sets 2 and 3.  The final simulation in Set 3 has a metallicity of 
10$^{-2} Z\subsun$, but unlike the others, excludes the effect of the CMB on the cooling of gas.  

We use the second implementation of the optically-thin metal cooling method of 
\citet{2008MNRAS.385.1443S}.  This methods uses tabulated cooling functions created with 
the photoionization software, Cloudy \citep{1998PASP..110..761F}, for all elements heavier 
than He, up to atomic number 30 (Zn).  A solar abundance pattern is used for the metals in 
all of the simulations.  Since we only follow the evolution of the collapsing 
clouds up to densities of $\sim$ 10$^{12}$ cm$^{-3}$, we neglect the cooling from dust and assume 
the presence of only gas-phase metals.  In \S\ref{sec:tau}, we calculate the optical depth 
in the collapsing cloud for the most important high density, gas-phase metal coolants and find 
that the optically thin assumption holds at all times during simulations.  
The H/He chemistry is followed explicitly during the 
simulation as in \citet{1997NewA....2..181A} and \citet{1997NewA....2..209A}, but including reactions that 
extend its validity to above 10$^{8}$ cm$^{-3}$.  See \citet{2008MNRAS.385.1443S} for a full discussion 
of this method.  We neglect cooling from HD, as \citet{2002ApJ...564...23B} found its contribution to 
be negligible in situations where the initial ionization is low, such as in this work.  
However, it has been pointed out by \citet{2008AIPC..990...25G} that when the initial ionization is high, 
the HD fraction can become enhanced to the point where HD cooling is important.  
Heating from H$_{2}$ formation is not included in 
the chemical network.  This may contribute a significant source of heat when H$_{2}$ formation via 
three-body reactions becomes important at $n \ga 10^{9}$ cm$^{-3}$ \citep{2005ApJ...626..627O}.  
However, we argue in \S\ref{sec:frag} that our results are robust despite this shortcoming.  
The metal cooling data was created with the Linux computer cluster Lion-xo which is 
operated by the High Performance Computing Group at The Pennsylvania State University.  

As in the precursor to this work \citep{2007ApJ...661L...5S}, we do not assume the existence of an 
ionizing UV background.  The Population III stars responsible for the enrichment of the gas in our 
simulations have died, and we assume any living Pop III stars are most likely too distant 
to contribute significant amounts of radiation.  A major difference between our simulations 
and those of \citet{2007ApJ...661L...5S} is the choice of H$_{2}$ cooling rates.  
We use the updated H$_{2}$ cooling rates of \citet{1998A&A...335..403G}, whereas 
\citet{2007ApJ...661L...5S} used the older rates of \citet{1984ApJ...280..465L}.  The 
rates of \citet{1998A&A...335..403G} are generally lower than those of 
\citet{1984ApJ...280..465L}, causing the gas to take longer to cool in the newer 
simulations.  As a result, the moment of runaway-collapse in our simulations 
is systematically delayed from those of \citet{2007ApJ...661L...5S}, when 
comparing the simulations in Set 1.

Each simulation is initialized at $z$ = 99 and runs until the point where at least one dense, 
prestellar core forms within a $\sim 5 \times$ 10$^{5}$ $M\subsun$ dark matter halo, located in 
the center of the simulation box.  The simulations are stopped when a maximum refinement 
of 24 levels below the top grid has been reached.  This corresponds to a maximum density of 
$\sim$ 10$^{11}$ cm$^{-3}$, or roughly 3 $\times$ 10$^{-13}$ g cm$^{-3}$.  
The simulations were run on DataStar, an IBM Power4 machine at the San Diego Supercomputing Center. 
A summary of the final state of each simulation is given in Table \ref{tab:sims}.

\section{Results} \label{sec:results}

\subsection{Critical Metallicities} \label{sec:Z_crit}

The gas phase critical metallicity, $Z_{cr}$, has been estimated analytically by calculating the 
chemical abundance required for the cooling time to equal the dynamical time at the stalling 
point for metal-free gas, $n$ $\sim$ 10$^{3-4}$ cm$^{-3}$ and $T$ $\sim$ 200 K.
\citet{2003Natur.425..812B} performed this exercise with C and O, but excluding cooling from H$_{2}$, and 
\citet{2006ApJ...643...26S} did so with C, O, Fe, Si, and including H$_{2}$.  
\citet{2006ApJ...643...26S} considered densities above and below the stalling 
point as well.  The general consensus from these studies is that $Z_{cr}$ $\approx$ 
10$^{-3.5}$ $Z\subsun$ at $n$ = 10$^{4}$ cm$^{-3}$.  

We observe a small amount of variance in the values of the density and temperature at 
which the temperature minimum occurs in our three metal-free runs.  Since the exact value of 
the critical metallicity depends on the precise conditions at the temperature minimum, i.e., 
temperature, total density, and H$_{2}$ density, we calculate the critical metallicity for 
each of the three runs separately.  To do this, we define the cooling time in the following 
way:
\begin{equation} \label{eqn:tcool}
t_{cool} = \frac{nkT}{(\gamma-1) [\Lambda_{H_{2}}^{\prime} + \Lambda_{metals}^{\prime}]},
\end{equation}
where $k$ is Boltzmann's constant, $\gamma$ is 5/3, and $\Lambda_{H_{2}}^{\prime}$ and 
$\Lambda_{metals}^{\prime}$ are the cooling rates from H$_{2}$ and the metals in units of 
[erg s$^{-1}$ cm$^{-3}$].  We use $\Lambda^{\prime}$ to avoid confusion with the term, $\Lambda$, 
which is often expressed in units of [erg s$^{-1}$ cm$^{3}$].  We then set Equation \ref{eqn:tcool} 
equal to the dynamical time,
\begin{equation} \label{eqn:tdyn}
t_{dyn} = \sqrt{\frac{3\pi}{16G\rho}},
\end{equation}
where $G$ is the gravitational constant and $\rho$ is the mass density.  We calculate $Z_{cr}$ 
for the conditions at the temperature minimum in each metal-free simulation using the H$_{2}$ 
cooling rates of \citet{1998A&A...335..403G} and the metal cooling rates from 
\citet{2008MNRAS.385.1443S}.  The results are shown in Table \ref{tab:zcrit}.  The systematic shift toward lower 
values of Z$_{cr}$ compared to previous calculations is most likely due to the fact that the 
minimum temperatures are slightly higher than 200 K, where both the H$_{2}$ and metal cooling rates 
are higher.  In addition, cooling from H$_{2}$ was not included in the calculation of 
\citet{2003Natur.425..812B}, meaning that more cooling from the metals would have been required.  
There is a correlation 
between $Z_{cr}$ and the collapse redshift of the simulation, with the highest value of $Z_{cr}$ 
coming from the highest collapse redshift, despite the lack of such a correlation for $n$, 
$n_{H_{2}}$, or $T$.  However, there is a correlation between the H$_{2}$ fraction and collapse 
redshift, which has been observed in the Pop III simulations of \citet{2007ApJ...654...66O}.  They 
find that the higher H$_{2}$ fractions result from the generally warmer gas in halos that collapse 
at higher redshifts, which is simply a function of the linear dependence upon redshift of the virial 
temperature.  This may be the dominant factor, 
but with only 3 data points it is unclear whether the observed trend is even significant.  In addition, 
the artificial nature of our initial conditions may make this finding inapplicable to the real world.

We see a trend with metallicity and collapse redshift that is similar to what was reported 
by \citet{2007ApJ...661L...5S}, where simulations with higher metallicities reach the 
runaway collapse phase earlier.  We define $\Delta t_{col}$ as the difference in time to 
runaway collapse between a simulation with non-zero metallicity and the metal-free run with 
the same initial conditions.  An increase in metallicity by 0.5 dex results in an increase in 
$\Delta t_{col}$ by a factor of approximately 1.3 to 4.  Similarly, when compared within a set, 
the simulations collapsing later have a higher number of total grids and grid cells in their 
final output, since the lower-density envelope gas has had more time to evolve and reach higher 
densities.

\begin{deluxetable}{lcccc}
  \tablecolumns{5}
  \tablewidth{0pt}
  \tablecaption{Critical Metallicities}
  \tablehead{
    \colhead{Run} &
    \colhead{$n$ [cm$^{-3}$]} &
    \colhead{$n_{H_{2}}$ [cm$^{-3}$]} &
    \colhead{$T$} & 
    \colhead{log($Z_{cr}$/$Z\subsun$)}}
  \startdata
  r1\_mf & 6.89$\times10^{3}$ & 3.42 & 283 & -4.08\\
  r2\_mf & 3.64$\times10^{3}$ & 1.86 & 214 & -3.90\\
  r3\_mf & 1.19$\times10^{4}$ & 6.53 & 260 & -3.85\\
  \enddata
  \tablecomments{Critical metallicities calculated for each metal-free simulation.  $n$, $n_{H_{2}}$, and $T$ 
    are the proper number density, H$_{2}$ number density, and temperature at the temperature minimum where H$_{2}$ 
    becomes thermalized in each of the metal-free simulations.  The final column is the log of the metallicity 
    required to equate the cooling time to the dynamical time for the conditions listed.}
  \label{tab:zcrit}
\end{deluxetable}

\subsection{Radial Profiles} \label{sec:radial}

\begin{figure*}
  \plotone{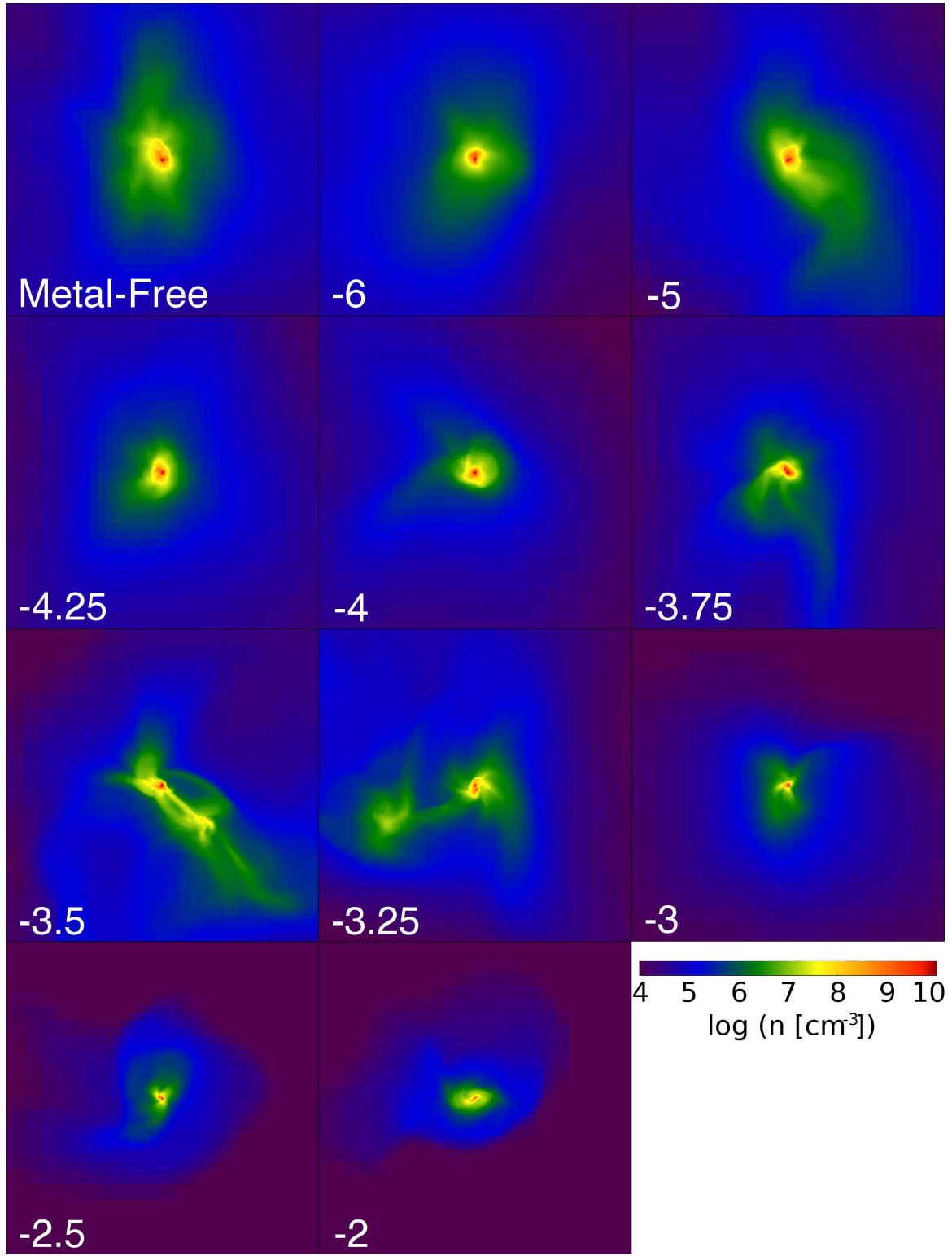}
  \caption{Projections of mass-weighed mean number density for the final output of all runs in 
    Set 1.  Each projection 
    is centered on the location of maximum density in the simulation box and has a width of 0.5 
    pc proper.  The labels in each panel indicate the log of the metallicity with respect to 
    solar for that run.  The images were made with the YT analysis toolkit 
    \cite[\texttt{yt.enzotools.org}]{SciPyProceedings_46}.
  } \label{fig:proj_08}
\end{figure*}

In Figure \ref{fig:proj_08}, we show projections of mass-weighted mean number density for the central 0.5 pc 
surrounding the point of maximum baryon density for the final output of all runs in Set 1.  For the runs with 
metallicities near or below $Z_{cr}$ (10$^{-4.08} Z\subsun$ for this set of simulations), the central cores 
appear quite round and show no clear signs of forming more than one object.  In the metallicity range from 
10$^{-3.75} Z\subsun$ to 10$^{-3.25} Z\subsun$, the cores appear increasing asymmetric, with at least one 
additional density maximum present.  However, at metallicities at or above 10$^{-3} Z\subsun$, the cores 
return to a more spherical shape with only a single density maximum.

\begin{figure*}
  \plotone{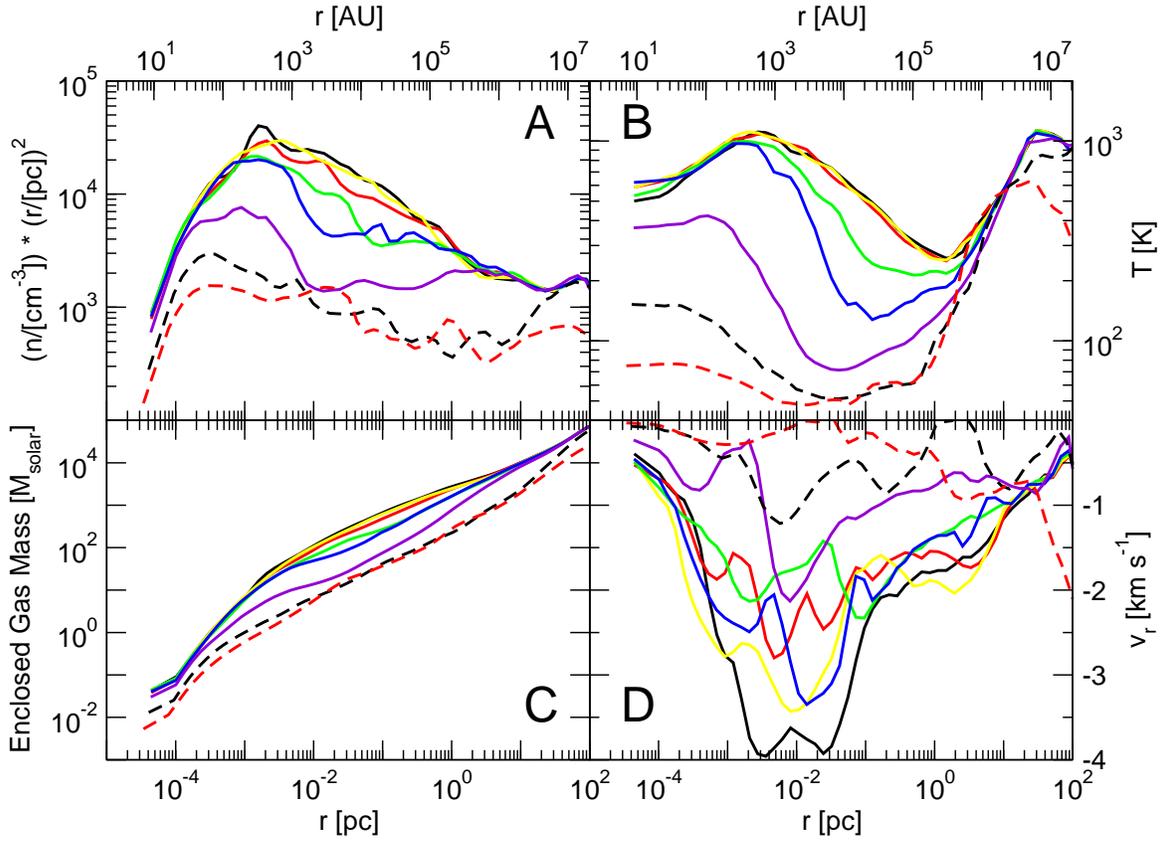}
  \caption{Spherically-averaged, mass-weighted quantities as a function of radius from 
  the point of maximum density for 7 of the 10 simulations in Set 1.  A: number density 
  normalized to an $r^{-2}$ power law; B: temperature; C: enclosed gas mass; D: 
  radial velocity.  In each panel, the metallicities are $Z$ = 0 (solid-black), $10^{-6} Z\subsun$ 
  (solid-red), $10^{-5} Z\subsun$ (yellow), $10^{-4} Z\subsun$ (green), $10^{-3.5} Z\subsun$ (blue), 
  $10^{-3} Z\subsun$ (purple), $10^{-2.5} Z\subsun$ (dashed-black), and $10^{-2} Z\subsun$ 
  (dashed-red).} \label{fig:radial_08}
\end{figure*}

In Figure \ref{fig:radial_08}, we plot spherically averaged, mass-weighted values of the density, 
temperature, enclosed mass, and radial velocity as a function of radius from the point of maximum 
density for all simulations in Set 1.  On large scales, the density profiles follow an r$^{-2.2}$ 
power law.  To highlight the difference in density between each run, we plot in panel A of 
Figure \ref{fig:radial_08} the value of ($n \times r^{2}$), instead of simply $n$.  We choose to 
scale the density by $r^{2}$ instead of $r^{2.2}$ because it is easier to glean the true density from 
the figure.  Over 
most of the plotted range, the run with the lowest metallicity has the highest density.  In the 
isothermal collapse model of \citet{1977ApJ...214..488S}, the accretion rate is proportional to 
the cube of the sound speed, or $T^{3/2}$.  Figure \ref{fig:radial_08}B shows that while 
isothermality does not really apply, there is a clear correlation between the temperature and 
density.  The runs with the highest metallicity, and subsequently the coldest gas, are the least 
dense.  In addition, within individual runs, an increase in the gas temperature is matched by an 
increase in the density.

The instantaneous accretion rate at a given radius is a function of 
the infall velocity at that position.  Although there are some exceptions, the correlation between 
temperature/sound speed and infall velocity, with higher temperatures/sound speeds corresponding to higher 
velocities, generally holds.  This was also found to be true by \citet{2007ApJ...654...66O} in their simulations of 
Pop III star formation.  We find that the inflow is roughly transonic throughout the entire density range, showing 
that even though the collapsing clouds are not isothermal spheres, their accretion rates are still largely 
regulated by the sound speed.

\subsection{Optical Depth} \label{sec:tau}

\begin{figure*}
  \plotone{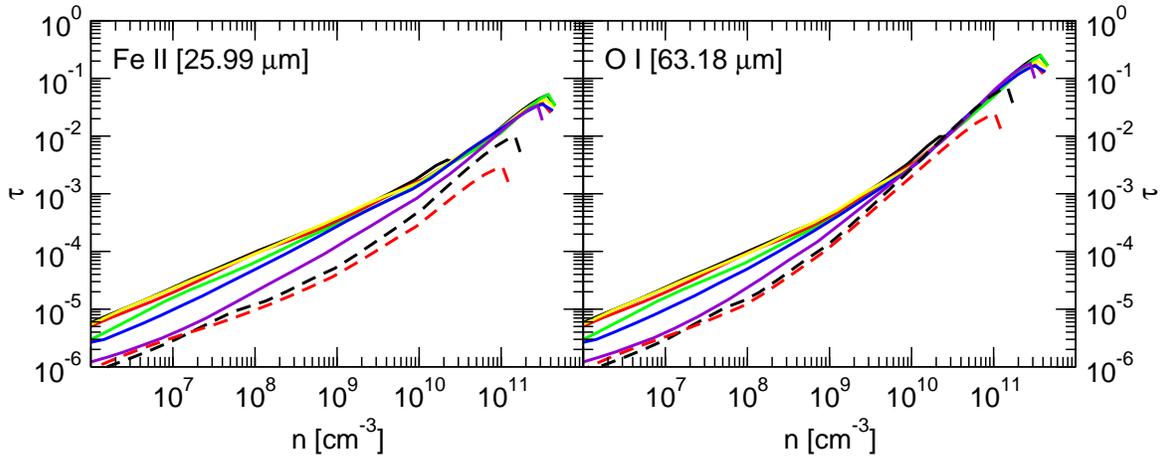}
  \caption{Optical depth for two of the most important high density, gas-phase coolants, Fe\textsc{ii} [25.99 $\mu m$] 
    (left) and O\textsc{i} [63.18 $\mu m$] (right), as a function of density for runs in Set 1.  The colors are the same as 
    in Figure \ref{fig:radial_08}.} \label{fig:tau_n_08}
\end{figure*}

The fragmentation of collapsing gas depends sensitively on its thermal evolution, which is effectively a 
measure of the radiative cooling properties of the gas as a function of density and temperature.  Since our 
radiative cooling method, described above, assumes optical thinness, the results of our simulations are only 
valid where $\tau \ll 1$.  The optically thin assumption begins to break down at densities $n \ga 10^{10}$ 
cm$^{-3}$ \citep{2005ApJ...626..627O}.  The optical depth at a frequency $\nu$ is expressed as
\begin{equation} \label{eqn:tau}
  \tau_{\nu} = \int  \kappa_{\nu} \: \rho  \: d\ell,
\end{equation}
where $\kappa_{\nu}$ is the opacity, $\rho$ is the mass density, and $d\ell$ is the distance traveled by a 
photon.  At a given density, the cloud has a characteristic size $r(\rho)$, shown in Figure \ref{fig:radial_08}A.  
If a photon emitted from a region with density $\rho$ is able to travel a distance $\sim r(\rho)$ without being 
re-absorbed, then it will not alter the thermal evolution of the cloud, and the assumption of optical thinness 
holds.  Using the Cloudy software, we calculate values of the absorption coefficient, $\alpha_{\nu} 
\equiv \kappa_{\nu} \: \rho$, as a function of density, metallicity, temperature, and frequency.  Equation 
\ref{eqn:tau} then takes the form
\begin{equation} \label{eqn:tau_sim}
  \tau_{\nu} = \alpha_{\nu}(\rho,Z,T) \: r(\rho).
\end{equation}
For metallicities $Z \la 10^{-2} Z\subsun$ and in the absence of dust, metal cooling is dominated by fine-structure 
transitions of O\textsc{i} and Fe\textsc{ii} \citep{2006ApJ...643...26S,2008MNRAS.385.1443S}.  In Figure 
\ref{fig:tau_n_08}, we plot the optical depth from Equation \ref{eqn:tau_sim} for the Fe\textsc{ii} line at 25.99 
$\mu$m and the O\textsc{i} line at 63.18 $\mu$m as a function of density using the 
spherically-averaged densities and temperatures shown in Figure \ref{fig:radial_08}.  
At low energies and in the absence of dust grains, the largest contributor to the opacity is free-free absorption, 
which is dominated by H for low metallicities.  As a result, the opacity in the energy range of interest is 
essentially independent of metallicity for metallicities less than solar.  When the metals become a considerable 
fraction of the total gas content, above $Z \sim 10 Z\subsun$ or so, the opacity at low energy for dust-free gas 
begins to grow with metallicity.  
In Figure \ref{fig:tau_n_08}, the apparent decrease in the optical depth at the 
highest metallicities is due to the flattening of the density profile at small radii.  As such, the 
final point for each curve in Figure \ref{fig:tau_n_08} should be ignored.  For values inside the point 
where the density profile flattens, it would be more reasonable to use the radius at which the density turns over 
for the calculation of $\tau$.  Fortunately, $\tau$ remains significantly less than 1 throughout all of our 
simulations, peaking at roughly 0.25 at the highest densities for the runs with the lowest metallicities.  The 
lower optical depth in the higher metallicity runs is due to the smaller characteristic size of the core, as seen 
in Figure \ref{fig:radial_08}A.  It should also be noted that some molecules are missing from the 
Cloudy dust-free chemistry network that could contribute to the opacity, such as TiO.  As such, the optical 
depths calculated could be slightly higher for the high metallicity simulations.  However, we show in 
\S\ref{sec:frag} that fragmentation in our simulations occurs well before this becomes a concern.

\subsection{Fragmentation} \label{sec:frag}

\begin{figure}
  \plotone{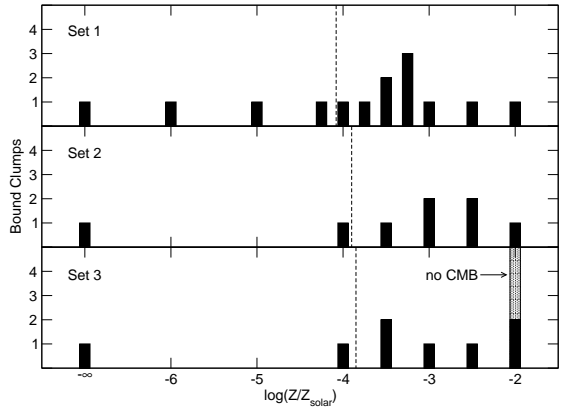}
  \caption{The number of bound clumps found within a sphere of radius 5 pc, centered on the 
  point of maximum density, as a function of metallicity for all of the simulations performed.  
  The grey bar in the bottom panel denotes the run where the temperature floor created by the CMB 
  was not included.  In each panel, the vertical dashed line represents the estimated value of 
  $Z_{cr}$ for each set from Table \ref{tab:zcrit}.} \label{fig:clumps}
\end{figure}

In order to quantify the degree of fragmentation within each run, we employ a clump finding 
algorithm to search for bound clumps within 5 pc of the density maximum.  As in 
\citet{1998ApJ...495..821T}, we define a clump, there referred to as a fragment, as ``the mass 
contained between a local density maximum and the lowest isodensity surface surrounding only that 
maximum,'' to quote that work.  We begin by identifying all grid cells within a sphere of radius 
5 pc, centered on the point of maximum density.  We then create density contours using all the 
cells within the sphere.  In density space, the first contour spans the entire range of density 
within the sphere, effectively creating one large contour.  The contour becomes the parent clump 
of all other clumps that will be found as the process continues.  On the second iteration, we 
create contours with the same maximum as before, but with the minimum increased by 1/4 dex.  If 
more than one contour exists, those groups of cells become child clumps of the group made by the 
previous iteration.  The process continues in a recursive fashion, creating groups of cells based 
on contours of increasing minimum density, identified only within the cells of the parent clump.  
The clump finding ends when minimum countour density has reached the constant maximum.  Effectively, 
we create a family tree of clumps, with the very first group as the trunk of the tree.  During 
the clump finding process, a child clump is only kept if it is gravitationally self-bound, or has 
children of its own that are bound.  When determining whether a clump is bound, we consider the 
thermal energy of the gas as well its kinetic energy with respect to the bulk center of motion.  

In Figure \ref{fig:clumps}, we plot a histogram of all the number of clumps found in each run within the 
5 pc sphere, as a function of the metallicity of the run.  For this plot, we only include bound child clumps 
with no children of their own.  When enlarging the radius of the sphere from 5 pc to 10 pc, no additional 
clumps were found in any of the runs.  Figure \ref{fig:clumps} confirms what is seen in Figure 
\ref{fig:proj_08}.  In all runs with metallicities below $Z_{cr}$, only a single bound clump is 
found.  As the metallicity increases, the number of clumps increases, then decreases back to only 
a single clump for the highest metallicities, with the exception of r3\_Z-2, which has 2 bound clumps.  
The range of metallicities where fragmentation occurs 
is consistent between Sets 1 and 3, but offset by 0.5 dex toward higher metallicities for Set 2.  
It is not clear what causes this offset, but the qualitative trend of increasing and then decreasing 
number of clumps exists in all 3 sets.  It is also worthwhile to note that runs r1\_Z-4, 
r1\_Z-3.75, and r2\_Z-3.5, while slightly above $Z_{cr}$, do not show fragmentation.  

\begin{deluxetable}{lcc}
  \tablecolumns{2}
  \tablewidth{0pt}
  \tablecaption{Fragmentation Properties}
  \tablehead{
    \colhead{Run} &
    \colhead{Clump Mass [$M\subsun$]} &
    \colhead{$n_{frag}$ [cm$^{-3}$]}}
  \startdata
  r1\_Z-3.5    & 54.4, 4.98 & 1.55 $\times 10^{7}$ \\
  r1\_Z-3.25   & 22.1       & 1.55 $\times 10^{6}$ \\
               & 5.46, 1.40 & 8.73 $\times 10^{6}$ \\
  r2\_Z-3      & 6.26, 1.89 & 2.75 $\times 10^{9}$ \\
  r2\_Z-2.5    & 95.6, 27.7 & 2.76 $\times 10^{4}$ \\
  r3\_Z-3.5    & 54.4, 4.98 & 1.55 $\times 10^{5}$ \\
  r3\_Z-2      & 289, 105   & 1.55 $\times 10^{4}$ \\
  r3\_Z-2noCMB & 11.7       & 2.76 $\times 10^{4}$ \\
               & 4.38, 4.06 & 4.91 $\times 10^{5}$ \\
               & 32.6, 0.49 & 8.73 $\times 10^{5}$ \\
  \enddata
  \tablecomments{Number density at which fragmentation occurs within simulations where multiple bound clumps 
    are identified and the masses of the clumps associated with that episode of fragmentation.  The fragmentation 
    density is taken to be the minimum density within a clump, taken from the 
    earliest data output where that clump is identified.}
  \label{tab:frag}
\end{deluxetable}

During the simulations, the Enzo code creates a snapshot of the entire box each time the maximum level 
of refinement increases.  This provides us with multiple data outputs as the central density increases 
during runaway collapse.  We ran the clump finder on all data outputs created during this period, searching 
for the first data output in which multiple clumps are found.  Within this output, we take the minimum 
density within the clump to be the density at which fragmentation occurred.  In Table \ref{tab:frag}, 
we list the fragmentation densities for all simulations in which multiple clumps were found.  In runs with 
2 clumps, the fragmentation density is the same for both clumps, since it represents the lowest density 
contour for which the two objects are separate.  In run r1\_Z-3.25, which has 3 clumps, the core initially 
fragmented into 2 clumps at $n = 1.55 \times 10^{6}$ cm$^{-3}$.  One data output later, one of those clumps 
fragmented again at $n = 8.73 \times 10^{6}$ cm$^{-3}$.  A similar thing occurred in run r3\_Z-2noCMB 
with 4 clumps forming initially and one of those fragmenting again later.

At densities of roughly 10$^{9}$ cm$^{-3}$, H$_{2}$ formation via three-body reactions begin to rapidly 
increase the H$_{2}$ fraction \citep{2002Sci...295...93A,2002ApJ...564...23B}.  In our simulations, we find 
the H$_{2}$ fraction to be roughly constant at $\sim 10^{-3}$ until the density reaches $\sim 10^{9}$ cm$^{-3}$.  
At densities of a few $\times$ 10$^{10}$ cm$^{-3}$, the H$_{2}$ fraction has passed 10\% and continues to rise.  
The onset of rapid H$_{2}$ formation occurs slightly earlier (within a factor of a few in density) in the runs 
with higher metallicity due to the $T^{-1}$ dependence of the three-body rate coefficients.  Our chemical network 
lacks heating from H$_{2}$ formation ($\sim$ 4.4 eV per reaction), which may significantly raise the temperature 
of the gas when three-body reactions become important \citep{2005ApJ...626..627O}.  We test this by calculating 
the ratio of the heating rate produced by H$_{2}$ formation to the total rate of cooling (without H$_{2}$ formation 
heating) for each of the simulations.  We use the rate coefficients $k_{4}$ (H + H + H $\rightarrow$ H$_{2}$ + H) 
and $k_{6}$ (H + H + H$_{2}$ $\rightarrow$ H$_{2}$ + H$_{2}$) from \citet{1983ApJ...271..632P} and assume an energy 
injection of 4.4 eV per reaction.  We also include a cooling term representing the reverse of the above two 
reactions, using coefficients $k_{5}$ and $k_{7}$ of \citet{1983ApJ...271..632P}.  For metallicities $Z \le 
10^{-3.5} Z\subsun$, the ratio of the H$_{2}$ formation heating to the total cooling is 0.1 at $n \sim 10^{9}$ 
cm$^{-3}$ and 1 at $n \sim 5 \times 10^{9}$ cm$^{-3}$.  For $Z = 10^{-3} Z\subsun$, this ratio is 0.1 at 
$n \sim 9 \times 10^{7}$ cm$^{-3}$ and 1 at $n \sim 6 \times 10^{8}$ cm$^{-3}$.  For $Z \ge 10^{-2.5} Z\subsun$, 
this ratio is 0.1 at $n \sim 3 \times 10^{7}$ cm$^{-3}$ and 1 at $n \sim 10^{8}$ cm$^{-3}$.  For run r3\_Z-2noCMB, 
which has the CMB removed, this ratio is 0.1 at $n \sim 10^{7}$ cm$^{-3}$ and 1 at $n \sim 10^{8}$ cm$^{-3}$.  
For simulations with equivalent metallicities, the above ratio varies maximally by a factor of a few due to 
minor differences in the temperature at a given density.  In all cases but one, fragmentation occurs well before 
H$_{2}$ formation heating becomes important.  The lone exception is run r3\_Z-3, where fragmentation occurs at 
$n \sim 3 \times 10^{9}$ cm$^{-3}$ and H$_{2}$ formation heating matches the total cooling at $n \sim 10^{9}$ 
cm$^{-3}$.  However, the core in this simulation does not fragment until it has already begun to reheat due to 
adiabatic compression, so it is not clear what effect the inclusion of this missing heating term would have had.

\begin{figure*}
  \plotone{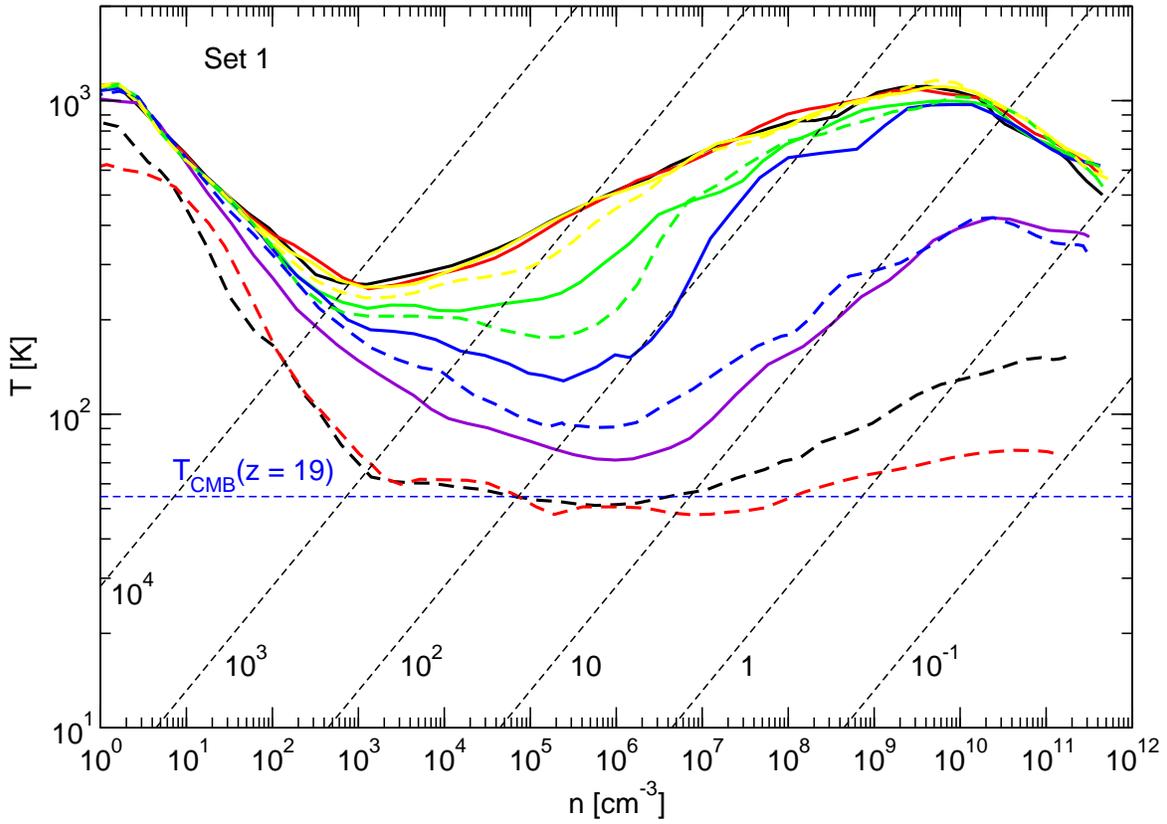}
  \caption{Mass-weighted, average temperature as a function of number density for all runs in 
    Set 1.  The colors are the same as in Figure \ref{fig:radial_08}, including the runs with 
    metallicities $Z$ = $10^{-4.25} Z\subsun$ (dashed-yellow), $10^{-3.75} Z\subsun$ (dashed-green), and 
    $10^{-3.25} Z\subsun$ (dashed-blue).  The thin, black, dashed lines indicate lines of constant 
    Jeans mass in $M\subsun$.  The horizontal, blue, dashed line denotes the temperature of the CMB at $z$ = 19, 
    the approximate redshift of collapse for runs r1\_Z-2.5 and r1\_Z-2.  The central cores in these two 
    runs were both able to cool to the temperature of the CMB.} \label{fig:T_n_08}
\end{figure*}

\begin{figure*}
  \plotone{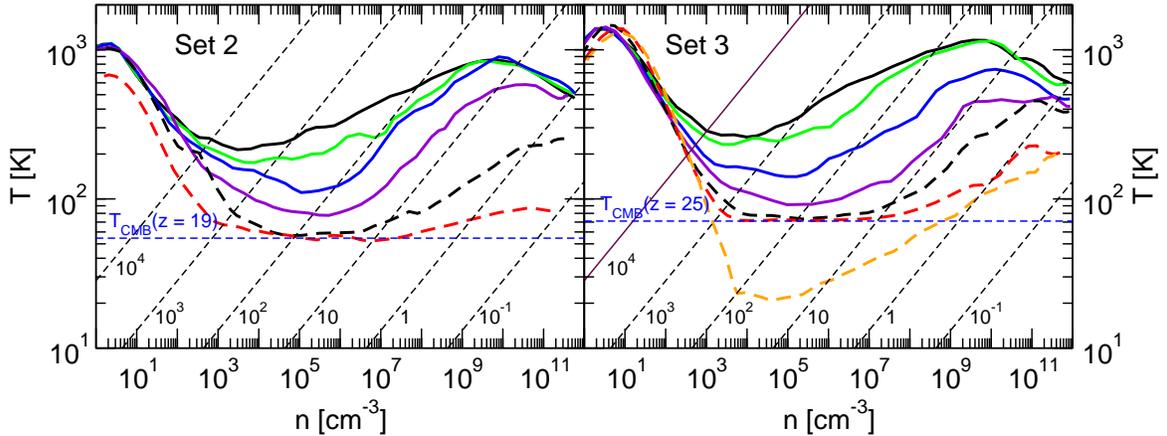}
  \caption{Mass-weighted, average temperature as a function of number density for all the runs in 
  Set 2 (left) and Set 3 (right).  For both panels, the metallicities are $Z$ = 0 (solid-black), 
  $10^{-4} Z\subsun$ (green), $10^{-3.5} Z\subsun$ (blue), $10^{-3} Z\subsun$ (purple), 
  $10^{-2.5} Z\subsun$ (dashed-black), and $10^{-2} Z\subsun$ (dashed-red).  In the bottom panel, 
  the dashed-orange line corresponds to run r3\_Z-2\_noCMB, with $Z = 10^{-2} Z\subsun$, but 
  with the CMB temperature floor removed.  All curves in Figures \ref{fig:T_n_08} and \ref{fig:T_n_real} 
  with the same colors refer to simulations with the same metallicities.  The thin, black, dashed lines indicate 
  lines of constant Jeans mass in $M\subsun$.  The horizontal, blue, dashed lines denote the temperature of the CMB at 
  $z$ = 19 (left) and 25 (right).} \label{fig:T_n_real}
\end{figure*}

In Figures \ref{fig:T_n_08} and \ref{fig:T_n_real}, we plot the number density vs. gas temperature 
for the final output of each simulation.  Due to the self-similar nature of the collapse, Figures 
\ref{fig:T_n_08} and \ref{fig:T_n_real} can also be used to understand the evolution of the central 
core throughout the collapse.  In all the runs with $Z < Z_{cr}$, the cooling is too low to prevent 
the temperature from rising at the H$_{2}$ thermalization density, $n$ $\sim$ 10$^{4}$ cm$^{-3}$.  
Therefore, the minimum fragmentation mass for these runs, set by the Jeans mass as the temperature 
minimum, is well over 1,000 $M\subsun$, which is nearly equivalent to the total enclosed mass.  
Even though runs r1\_Z-4 and r1\_Z-3.75 are above $Z_{cr}$, the additional cooling provided by the 
metals is not sufficient to significantly lower the minimum fragmentation mass.  For runs 
r1\_Z-3.5 and r1\_Z-3.25, the more efficient cooling lowers the minimum fragmentation mass to just 
over 100 $M\subsun$, which is approximately a factor of a few lower than the total mass within 1 pc.  

For the runs with the highest metallicities, as in runs r1\_Z-2.5 and r1\_Z-2, the gas cools 
all the way to the temperature of the CMB.  The cooling proceeds so efficiently that the gas has 
not had sufficient time to reach high densities before hitting the temperature floor of the CMB.  
Fragmentation can only continue as long as the temperature decreases with increasing density 
\citep{1985MNRAS.214..379L,2005MNRAS.359..211L}.  Although the temperature decreases slightly 
in runs r1\_Z-2.5 and r1\_Z-2 for densities greater than 10$^{3}$ cm$^{-3}$, the temperature 
minimum is effectively at $n$ = 10$^{3}$ cm$^{-3}$, where the gas reaches the CMB temperature.  
Near the CMB temperature, the value of the cooling rate, $\Lambda$, effectively becomes 
($\Lambda(T) - \Lambda(T_{CMB})$).  Therefore, when the gas reaches the CMB temperature, the cooling 
rate drops to zero, and the cooling time becomes infinite.  The gas cloud becomes extremely 
thermally stable, preventing further fragmentation.

\begin{figure*}
  \plotone{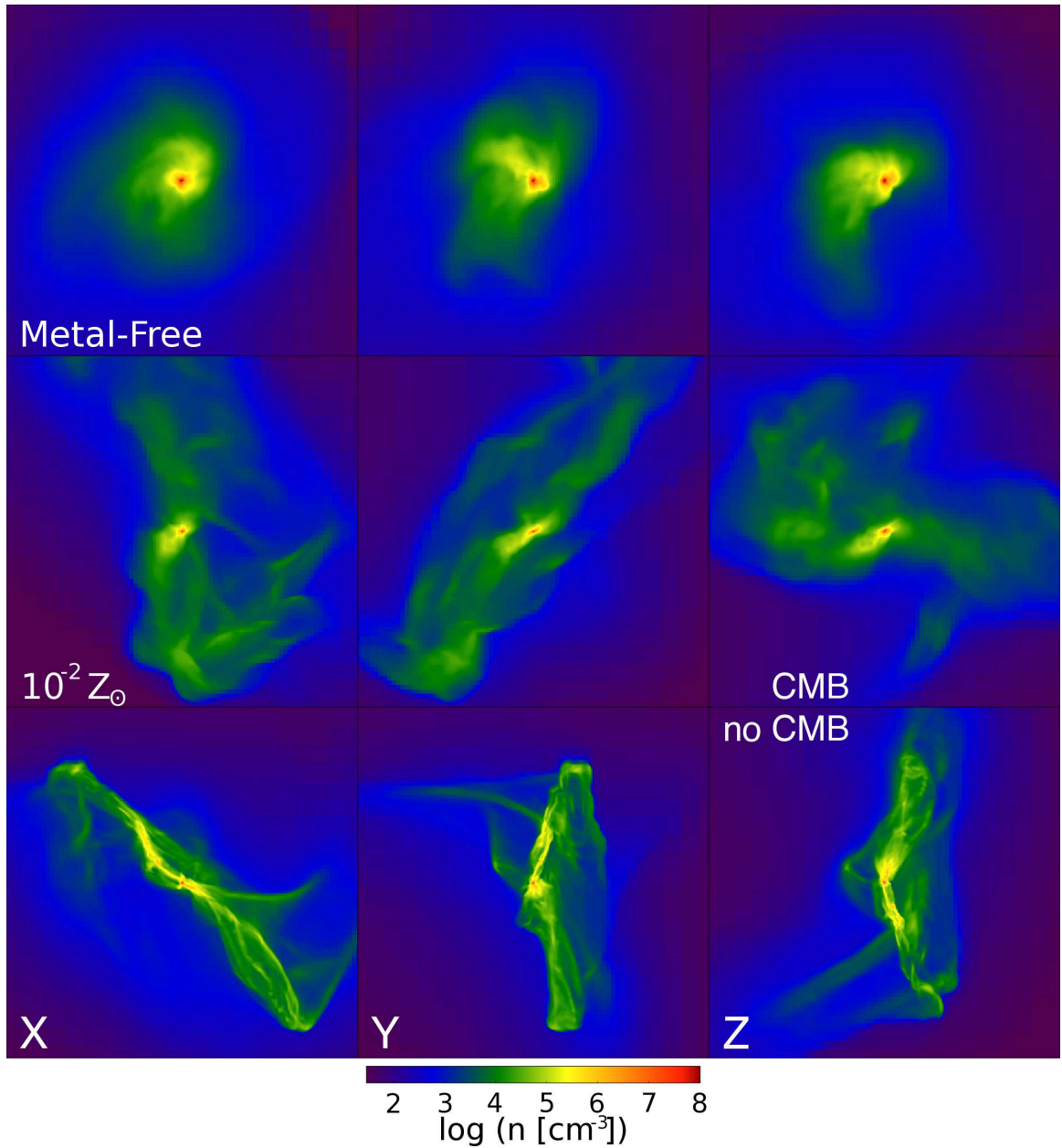}
  \caption{Projections of mass-weighted mean number density along the x (left), y (center), and z (right) axes for the 
    final output of runs r3\_mf with zero-metallicity (top), r3\_Z-2 with $Z$ = $10^{-2} Z\subsun$ (middle), and 
    r3\_-2noCMB with $Z$ = $10^{-2} Z\subsun$ and the CMB temperature floor removed (bottom).  Each projection 
    is centered on the location of maximum density in the simulation box and has a width of 5 
    pc proper.  The images were made with the YT analysis toolkit 
    \cite[\texttt{yt.enzotools.org}]{SciPyProceedings_46}.} \label{fig:proj_cmb}
\end{figure*}

\begin{figure*}
  \plotone{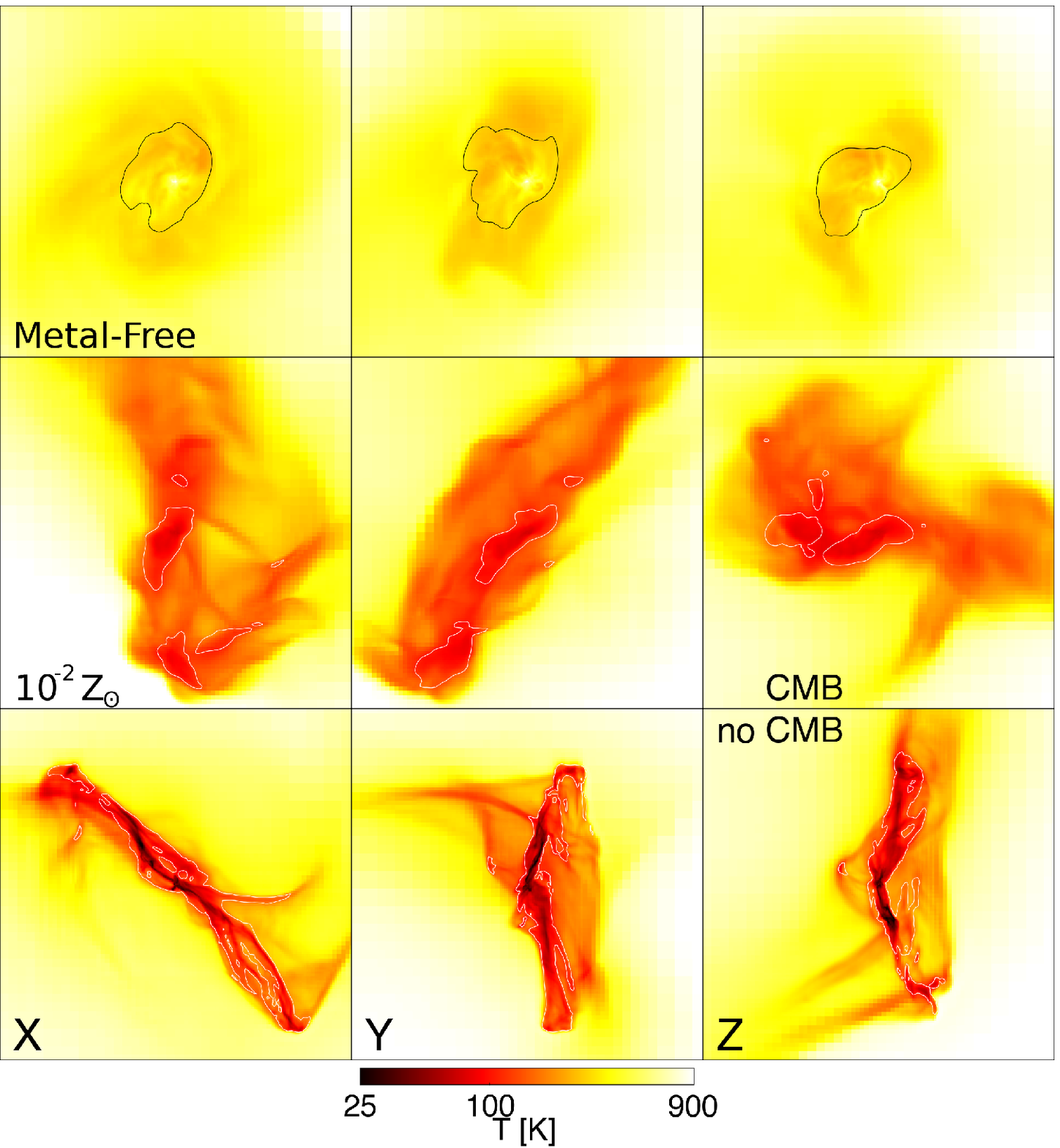}
  \caption{Projections of mass-weighted mean temperature density along the x (left), y (center), and z (right) axes 
    for the final output of runs r3\_mf with zero-metallicity (top), r3\_Z-2 with $Z$ = $10^{-2} Z\subsun$ (middle), 
    and r3\_-2noCMB with $Z$ = $10^{-2} Z\subsun$ and the CMB temperature floor removed (bottom).  Each projection 
    is centered on the location of maximum density in the simulation box and has a width of 5 pc proper.  The solid 
    lines show contours of projected mean number density of 10$^{4}$ cm$^{-3}$.  The images were made with the YT 
    analysis toolkit \cite[\texttt{yt.enzotools.org}]{SciPyProceedings_46}.} \label{fig:proj_cmb_temp}
\end{figure*}

To verify that the CMB is indeed suppressing 
fragmentation, we run one simulation, r3\_Z-2\_noCMB, with the CMB temperature floor removed.  We 
choose initial conditions Set 3 for this exercise since it has the highest CMB temperature at the 
redshift of collapse and 
should therefore show the greatest contrast with the CMB removed.  In Figure \ref{fig:proj_cmb}, 
we show mass-weighted mean number density projections of the central 5 pc for runs, r3\_mf, r3\_Z-2, and 
r3\_Z-2\_noCMB.  Run r3\_Z-2 has a much clumpier structure than its metal-free counterpart, even with the CMB 
temperature floor present.  However, when the temperature floor is removed, the gas is able to 
collapse into a long, thin filament with far more small-scale structure.  As shown in Figure 
\ref{fig:clumps}, we find the most bound clumps in this run (4 within 1 pc of the density peak 
and 1 more within 5 pc).  In Figure \ref{fig:proj_cmb_temp}, 
we show mass-weighted mean temperature projections for the same runs as Figure \ref{fig:proj_cmb}.  We 
overlay contours of projected mean number density of 10$^{4}$ cm$^{-3}$.  Figure \ref{fig:T_n_real} shows that 
this is the approximate density at which the cloud in run r3\_Z-2 first reaches the CMB temperature.  
In run r3\_Z-2noCMB, it took $\sim$400,000 years for the central density to increase from 10$^{4}$ cm$^{-3}$ 
to 10$^{5}$ cm$^{-3}$, which is similar to the timescale in a dynamical collapse.  In run r3\_Z-2, the equivalent 
change in density took $\sim$1.9 million years, indicating that cooling to the CMB temperature has indeed 
ended free-fall collapse.  

In the projections of run r3\_Z-2 in 
Figure \ref{fig:proj_cmb_temp}, the 2 largest contours roughly represent the 2 bound clumps found within the 5 pc 
radius sphere.  The gas within the clumps has a very uniform temperature, as its cooling has been 
abruptly halted at the CMB temperature ($\sim$71 K).  The 2 clumps have masses of roughly 100 and 300 
$M\subsun$.  It is unlikely that they will fragment further, since the Jeans mass of each clump is nearly 
equivalent to its total mass.  At a temperature of 71 K, sound waves will travel $\sim$1.7 pc in the 1.9 Myr 
required for any additional fragments to increase in density by an order of magnitude once they have cooled to 
the CMB temperature.  Any additional fragments that might possibly condense out of lower densitiy gas must be 
approximately this large in order to collapse, suggesting they will also be very massive.  It is interesting 
to note that the two bound clumps in the simulation are also about this size.

In contrast, the gas inside the contours of run 
r3\_Z-2noCMB shows significantly more structure in temperature.  The cold knots in run r3\_Z-2noCMB seen in 
Figure \ref{fig:proj_cmb_temp} correspond to the high density regions seen in Figure \ref{fig:proj_cmb}.  4 of 
the 5 bound clumps are Jeans unstable, with masses of 4, 4, 12, and 33 $M\subsun$.  The fifth, with 
$M \sim 0.5 M\subsun$, is approximately 1/6 of its Jeans mass.  The most massive clump is about 5 times more 
massive than its Jeans mass.

\section{Discussion} \label{sec:discussion}

We have shown that fragmentation occurs within a collapsing cloud when the 
metallicity is above the critical metallicity.  
The exact value of the critical metallicity required to prevent an increase in temperature at the 
stalling point of H$_{2}$ varies slightly from halo to halo.  Within our three sets of initial 
conditions, the values of $Z_{cr}$ are correlated to the collapse redshift of the metal-free runs, 
with the highest Z$_{cr}$ corresponding to the highest redshift.  However, it is unclear whether 
this is significant.  If the metallicity is only marginally higher 
than $Z_{cr}$, fragmentation is unlikely to occur, as the increase in temperature at the H$_{2}$ 
stalling point is only delayed momentarily.  Thus, it is unlikely that a sharp transition in 
star formation mode occurs at just the moment when the critical metallicity is reached.

Fragmentation is suppressed when the metallicity is 
high enough such that the gas is able to cool to the temperature of the CMB when the central 
density is still relatively low.  
We confirm that the CMB is responsible for the observed 
suppression of fragmentation by running an identical simulation without the CMB.  In this simulation 
where the CMB is absent, we find more bound clumps than in any other of the runs in this study.  
2 bound clumps were found in run r3\_Z-2, where the gas was able 
to cool to the CMB temperature.  However, both of these clumps were quite massive ($M \ga 100 M\subsun$), 
and we showed in \S\ref{sec:frag} that is is unlikely that they will fragment into smaller objects.  
We observe a small amount of variance in the metallicity range in which fragmentation occurs that 
does not appear to be related to the CMB.  Just as the exact value of $Z_{cr}$ seems to vary from 
halo to halo, we suspect that the range of metallicities where fragmentation occurs will also be 
influenced by the individual properties of a halo as well as its particular evolution.

The mass scale of collapsing clumps can be estimated from the Jeans mass at the end of the cooling 
phase.  This implies the existence of three distinct metallicity regimes for star formation.  In 
the first regime, $Z$ $\la$ $Z_{cr}$, which we refer to as the `primordial' mode, metals do not 
provide enough additional cooling to allow the gas 
temperature to continue to decrease monotonically with increasing density when the core reaches the 
H$_{2}$ thermalization density.  In this case, the collapse proceeds in a similar way to the 
metal-free scenario, resulting in the formation of a single, massive object.  There is also the 
potential for stars forming in this mode to be somewhat less massive than the very first stars.  
The fragmentation mass scale for extremely low metallicity gas ($Z \ll Z_{cr}$) may be lowered though 
compression by shocks from Pop III supernovae and enhanced cooling from HD in relic HII regions 
\citep{2003ApJ...586....1M,2006MNRAS.366..247J}.  

At the other extreme, we define 
$Z_{CMB}$ as the metallicity at which the gas can cool to the CMB temperature.  When $Z \gg Z_{CMB}$, 
the cloud-core will efficiently cool to the temperature of the CMB when the central 
density is still relatively low.  In this scenario, fragmentation is limited by cooling rapidly to 
the CMB temperature, as the mass scale is determined by the Jeans mass at the density when the core 
first reaches the CMB temperature.  We refer to this as the CMB-regulated star formation mode, similar 
to \citet{2007ApJ...664L..63T}.  As 
fragmentation is severely limited in this mode, these stars will most likely be more massive on average 
than the characteristic mass of stars forming today.

Finally, our simulations have shown that there 
exists a special range in metallicity, $Z_{cr} \le Z < Z_{CMB}$, where the core does not reheat 
at the metal-free stalling point, but also cannot cool all the way to the CMB temperature.  The minimum 
temperature is set only by the balance of radiative cooling and adiabatic heating.  The mass scale is 
not regulated externally by the CMB, but rather internally by the metallicity-dependent gas-cooling.  
Hence, we term this the metallicity-regulated star formation mode.  This mode produces the lowest mass 
stars of the three modes mentioned.

The CMB-regulated star formation mode creates a means by which a higher number of massive stars are 
formed in very early universe, when the CMB temperature was much higher.  As the universe evolves, the 
CMB temperature will slowly decrease, which will increase the metallicity required to reach the CMB 
temperature, referred to here as $Z_{CMB}$.  The decrease in the CMB temperature also means that the 
fragmentation mass scale will be lower at the point where the gas reaches the temperature floor.  Thus, 
the characteristic mass of stars produced by the CMB-regulated mode will slowly decrease with time.  
This behavior is in agreement with the model of an IMF that evolves with redshift formulated by 
\citet{1998MNRAS.301..569L}.  As the metallicity threshhold for the 
CMB-regulated mode advances to higher metallicity, the range of operation of the metallicity-regulated 
mode will extend to take its place.  We lack sufficient data in this study to predict the evolution of 
$Z_{CMB}$ with redshift.  However, in a paper to follow, we will map out the evolution of $Z_{CMB}$ vs. 
$z$ with additional simulations collapsing as much lower redshifts.  
Observations of nearby star-forming clouds show that the minimum 
achievable temperature in the local universe is roughly 10 K (e.g., \citet{1999ARA&A..37..311E}).  This 
implied that the CMB-regulated star formation mode is in operation up to $z \sim 2.7$, at the absolute 
latest.  
A growing amount of evidence has been presented that the stellar IMF evolves with redshift (e.g., 
\citet{2007MNRAS.379..985F,2007ApJ...664L..63T,2008ApJ...674...29V,2008MNRAS.385..147D,
2008MNRAS.385..687W}).  Interestingly, 
\citet{2008ApJ...674...29V}, \citet{2008MNRAS.385..147D}, and \citet{2008MNRAS.385..687W} report 
evidence from high redshift that the 
IMF may deviate from the standard Salpeter IMF, favoring higher mass stars for $z \ga$ 2 - 4.

If dust is present in the very early universe, this would extend the range of the metallicity-regulated 
star formation mode to metallicities as low as $Z \sim 10^{-5.5} Z\subsun$ 
\citep{2005ApJ...626..627O,2006MNRAS.369.1437S,2006ApJ...642L..61T,2008ApJ...672..757C}.  The existence 
of low mass, hyper metal poor stars, HE0107-5240 \citep{2002Natur.419..904C}, and HE1327-2326 
\citep{2005Natur.434..871F}, both with [Fe/H] $<$ -5, may provide evidence of this.  Both of these stars 
show extremely enhanced C and O abundances, which would make their effective metallicities (in terms 
of the radiative cooling ability of gas with that abundance) much higher.  However, it is pointed out 
by \citet{2007ApJ...665.1361T,2007ApJ...664L..63T}, that the abundance patterns of these stars are best 
recreated by a scenario in which the C and O enhancement comes via binary mass transfer from an 
intermediate mass AGB star, meaning the stars are truly metal poor.  In that case, these two low-mass 
stars would likely require dust in order to form at such low metallicity.  \citet{2007ApJ...664L..63T} also 
shows that such a high fraction of carbon enhanced metal poor stars (CEMPs) requires a higher than normal 
fraction of more massive stars that go through the AGB phase.  
\citet{2007ApJ...664L..63T} claims that the evolution of CEMP fraction with metallicity (with a higher CEMP fraction 
at lower metallicity) already shows the influence of the CMB on the IMF.  If a CEMP star 
and its binary companion were formed from gas at the same metallicity, there would have to be process 
at work that would prevent the dust cooling fragmentation from forming only low mass stars.  The star 
formation models of \citet{2005ApJ...626..627O} do not indicate that the dust cooling phase that induces 
low mass fragmentation is able to reach the CMB temperature for $Z < 10^{-4} Z\subsun$.  This may simply 
imply that metal mixing from the first supernovae is higly heterogeneous, allowing stars to form 
simultaneously with largely different abundances.

In this work, we study only gas clouds with solar abundance patterns.  Most likely, the first metals in the 
universe will not have solar abundance patterns.  However, from the perspective of simulating metal-enriched 
star formation, it is not the specific elemental abundances of a gas cloud that are important, but rather the 
total cooling rate produced by the gas.  Therefore, given that it is the sum of the metals that is important, and 
not the abundance pattern, the results of this work are robust in spite of the fact that the abundance patterns used 
are likely to be incorrect.

There appears to be some discrepancy between our results and those of \citet{2007ApJ...660.1332J}, who see 
no evidence of fragmentation induced by gas-phase metal cooling.  This is potentially resolved by 
the fact that in that work, the gas collapse is only strictly followed up to densities of 
5$\times10^{2}$ cm$^{-3}$ before sink particles are created.  Figures \ref{fig:T_n_08} and 
\ref{fig:T_n_real} show very little difference in the thermal structure of the gas for densities 
below 5$\times10^{2}$ cm$^{-3}$.  Additionally, our clump finding algorithm found only a single bound 
clump within every simulation when the analysis was performed on data outputs that were made when 
the maximum density was only $\sim10^{3}$ cm$^{-3}$.  Multiple clumps were only found when the clouds 
had reached somewhat higher densities.  Finally, the simulations in this work began with cold, neutral 
gas, whereas their simulations began with hot, ionized gas.  Had their simulations been run to higher 
densities, any dissimilarities might also be due to using different initial conditions.

\section{Conclusion} \label{sec:conclusion}

We have performed a series of high resolution simulations of metal-enriched star formation using 
cosmological, Pop III style initial conditions, and assuming fully homogeneous metal enrichment.  
We have shown that our results apply to more than a single star forming region by using 3 different 
sets of initial conditions with identical cosmological parameters and resolution, but with 3 unique 
random seeds with which to create the initial perturbations in the density and velocity fields.  From 
the results of these simulations, the main conclusions of this work are:

1. Fragmentation does not occur when the metallicity is only slightly above $Z_{cr}$, since this only 
leads to a small delay in the onset of the loitering phase that is brought on by a decrease in the efficiency of 
H$_{2}$ cooling.  The density at which the temperature begins to increase with increasing density is only marginally 
higher than in the metal-free case, and therefore does not lead to a signicant lowering of the minimum Jeans 
mass.  Within our simulations, $Z_{cr}$ is roughly 10$^{-3.9} Z\subsun$.  
We find that fragmentation does not occur until the metallicity is roughly 0.5 dex above $Z_{cr}$.

2. Fragmentation is suppressed when the metallicity is high enough such that the gas is able to cool to the 
temperature of the CMB when the density of the collapsing cloud is still relatively low ($n \sim 10^{4}$ cm$^{-3}$).  
The Jeans mass at the density and temperature at which the cloud first reaches the CMB temperature sets the minimum 
fragmentation mass within the cloud, and as such only massive clumps are able to form.

3. Metal-enriched star formation occurs in three modes that are separated by two metallicity thresholds, $Z_{cr}$ 
and $Z_{CMB}$.  $Z_{cr}$ is the conventional critical metallicity and $Z_{CMB}$ is the metallicity at which the 
gas is able to cool to the temperature of the CMB at a given redshift.  At the approximate collapse redshifts 
of our simulations, $z \sim 20$, $Z_{CMB}$ is between 10$^{-3}$ and 10$^{-2.5} Z\subsun$.  The three modes of star 
formation are:
\begin{itemize}
\item The primordial mode ($Z < Z_{cr}$) - The additional cooling provided by the metals is not enough to 
significantly alter the thermal structure of the cloud relative to the metal-free case.  No fragmentation occurs 
and the star will have a mass similar to a Pop III star.  
\item The metallicity-regulated mode ($Z_{cr} < Z < Z_{CMB}$) - Metal cooling is high enough to allow the cloud to 
continue to cool past the metal-free loitering phase, but not high enough to cool it to the temperature of the CMB.  
The minimum fragmentation mass is set at lower temperatures and higher densities than in the primordial case and 
fragmentation into multiple objects occurs.  Based on the masses of clumps formed within our simulations, stars 
forming in this mode could have masses of the order of a few $M\subsun$ or less.
\item The CMB-regulated mode ($Z > Z_{CMB}$) - Fragmentation is suppressed when the cloud is able to cool to the CMB 
temperature, as is described in point 2 of the conclusions.  At minimum, stars forming in the CMB-regulated mode 
will be more massive than stars forming in the metallicity-regulated mode.  $Z_{CMB}$ will increase as the CMB 
temperature lowers with time.  As such, the masses of stars forming in the CMB-regulated mode will slowly decrease.  
This mode will vanish altogether when the CMB temperature reaches the observed minimum temperature of nearby 
molecular clouds ($T \sim$ 10 K at $z \sim$ 2.7).
\end{itemize}

4. By demonstrating that the CMB can suppress fragmentation, we have provided a key conceptual piece 
CMB-IMF hypothesis \citep{2007ApJ...664L..63T}.  As pointed out by \citet{2007ApJ...664L..63T}, an 
IMF that evolves with redshift, producing more massive stars in the past, has consequences that may already 
be testable by observations.

\acknowledgments
We are extremely grateful to an anonymous referee who provided comments that significantly 
strengthened the arguments presented and helped to improve the manuscript a great deal.  
BDS would like to thank Jason Tumlinson and Simon Glover for extremely insightful discussions, as well as 
Gary Ferland and Peter van Hoof for their assistance with the Cloudy code.  
BDS and SS thank the Aspen Center for 
Physics for their hospitality during the 2008 Aspen Winter Conference On Astrophysics.  
This work was made possible by Hubble Space Telescope Theory Grant HST-AR-10978.01.  BWO and BDS carried 
out portions of this work under the auspices of the National Nuclear Security Administration of the U.S. 
Department of Energy at Los Alamos National Laboratory under Contract No. DE-AC52-06NA25396.  
BDS was supported at the University of Colorado, Boulder by NASA Theory grant NNX07AG77G.  BWO was 
supported by a LANL Director's Postdoctoral Fellowship (DOE LDRD grant 20051325PRD4).  MJT performed this 
work under the auspices of the Department of Energy at the Stanford Linear Accelerator Center in the Kavli 
Institute for Particle Astrophysics and Cosmology under Contract No. DE-AC02-76SF00515.  
The simulations were performed at SDSC with computing 
time provided by NRAC allocations MCA98N020 and TG-AST070010N.

\end{document}